\documentclass[acmsmall]{acmart}
\AtBeginDocument{%
  }

\setcopyright{cc}
\setcctype{by}
\acmDOI{10.1145/3832149}
\acmYear{2026}
\acmJournal{PACMSE}
\acmVolume{3}
\acmNumber{ISSTA}
\acmArticle{ISSTA058}
\acmMonth{10}
\acmSubmissionID{issta26main-p483-p}
\received{2026-01-30}
\received[accepted]{2026-06-25}

\usepackage{algorithm}
\usepackage{algpseudocode}
\usepackage{textcomp}
\usepackage{xcolor}
\usepackage{enumitem}
\usepackage{booktabs}
\usepackage{tabularx}
\usepackage{multirow}
\usepackage{array}
\usepackage{graphicx}
\usepackage{subcaption}
\usepackage{adjustbox}
\usepackage{makecell}
\usepackage[most]{tcolorbox}
\usepackage[switch]{lineno} 
\usepackage{url}
\usepackage{textcomp}
\usepackage[normalem]{ulem} 

\newcommand{\add}[1]{{#1}} 
\newcommand{\note}[1]{} 
\newcommand{\del}[1]{}
\definecolor{revcolor}{RGB}{0,92,230} 
\definecolor{bbb}{RGB}{0,0,255} 

\newcommand{\swebench}{\textit{SWE-Bench}}
\newcommand{\mswebench}{\textit{Multi-SWE-Bench}}
\newcommand{\swev}{\textit{SWE-Bench-Verified}}
\newcommand{\mswef}{\textit{Multi-SWE-Bench-Flash}}

\newcommand{\obsmask}{\texttt{ObsMask}}
\newcommand{\original}{\texttt{Original}}
\newcommand{\attncompress}{\texttt{AttnCompress}}
\newcommand{\agentdiet}{\texttt{AgentDiet}}
\newcommand{\lingua}{\texttt{Lingua}}
\newcommand{\summary}{\texttt{LLMSummary}}
\newcommand{\random}{\texttt{Random}}
\newcommand{\sliding}{\texttt{SlidingWindow}}

\newcolumntype{C}[1]{>{\RaggedRight\arraybackslash}p{#1}}
\NewDocumentCommand{\mybox}{ m m }{%
    \begin{tcolorbox}[
        enhanced,
        colback=gray!10,
        frame hidden,
        boxrule=0pt,
        arc=2mm,
        boxsep=2pt,
        left=6pt,
        right=6pt,
        top=6pt,
        bottom=2pt,
        title=\textbf{#1}, 
        coltitle=black,
        fonttitle=\bfseries\sffamily,
        attach boxed title to top left={yshift=-2mm, xshift=3mm},
        boxed title style={colback=white,boxrule=0pt,frame hidden,colback=gray!40}
    ]
    #2 
    \end{tcolorbox}
}

\begin{document}

\title{AttnCompress: Dynamic Attention-Guided Trajectory Compression for Software Engineering Agents}

\author{Zhengran Zeng}
\authornote{Both authors contributed equally to this research.}
\orcid{0009-0009-8422-4522}
\affiliation{%
  \institution{Peking University}
  \city{Beijing}
  \country{China}
}
\email{zhengranzeng@stu.pku.edu.cn}

\author{Yixin Li}
\authornotemark[1]
\orcid{0009-0005-0492-2234}
\affiliation{%
  \institution{Peking University}
  \city{Beijing}
  \country{China}
}
\email{leason\_lyx@stu.pku.edu.cn}

\author{Rui Xie}
\authornote{Those authors are the corresponding authors.}
\orcid{0000-0002-1756-7746}
\affiliation{%
  \institution{Peking University}
  \city{Beijing}
  \country{China}
}
\email{ruixie@pku.edu.cn}

\author{Wei Ye}
\authornotemark[2]
\orcid{0000-0002-9331-4716}
\affiliation{%
  \institution{Peking University}
  \city{Beijing}
  \country{China}
}
\email{wye@pku.edu.cn}

\author{Shikun Zhang}
\authornotemark[2]
\orcid{0000-0002-8576-2674}
\affiliation{%
  \institution{Peking University}
  \city{Beijing}
  \country{China}
}
\email{zhangsk@pku.edu.cn}

\begin{abstract}

The transition from human-centric assistance to Autonomous Software Engineering (ASE) agents has enabled the resolution of complex real-world SE tasks. However, the trial-and-error nature of these agents generates lengthy interaction trajectories, creating severe bottlenecks in terms of context window limits and cost. While context compression offers a potential remedy, prior approaches suffer from static pruning strategies and granularity mismatches, often failing to preserve the semantic dependencies and syntactic details crucial for SE tasks.
To strictly preserve critical task evidence while reducing context length, we introduce \attncompress{}, a dynamic attention-guided trajectory compression framework. 
Unlike existing approaches, \attncompress{} bridges the gap between semantic integrity and dynamic adaptability through three key mechanisms: (1) structure-aware segmentation via perplexity (PPL) spikes to preserve the syntactic structure of code and logs; (2) relevance estimation using proxy attention weights to quantify the precise relevance of historical blocks to the agent's current reasoning; and (3) a dynamic rolling window to re-evaluate and recall historical context as the task evolves. 
Extensive evaluation on \swev{} and \mswebench{} demonstrates that \attncompress{} achieves a pass rate of 53.17\%, outperforming prior state-of-the-art baselines while reducing token consumption by 21.6\% and total costs by 33.6\%. The framework proves to be model-agnostic and generalizes effectively across diverse programming languages.
\end{abstract}



\maketitle

\section{Introduction}
\label{sec:introduction}

The advent of Large Language Models (LLMs) has catalyzed a paradigm shift in software engineering (SE), transitioning from human-centric assistance to Autonomous Software Engineering (ASE) agents~\cite{LLMBasedAgentsforSE_Survey,CodeLLMandAgentsSurvey}. Capable of perceiving, reasoning, and acting, these agents have demonstrated remarkable potential in resolving complex real-world GitHub issues, as evidenced by benchmarks like \swebench{}~\cite{SWE-bench} and its extensions~\cite{Multi-SWE-bench,SWEbenchMultilingual,SWE-bench-Multimodal}. More broadly, SE agents have also been applied to a wide range of software engineering tasks that require sustained interaction with large codebases, including repository-level code understanding, feature implementation and test generation~\cite{FromLLMstoLLM-basedAgents,SWE-Compass}. Unlike simple Q\&A tasks, SE agents operate through long-horizon interactions, engaging in iterative ``trial-and-error'' workflows involving codebase analysis, file editing, test execution, and debugging~\cite{AStudyofTrajectories}. This process inevitably generates lengthy interaction trajectories.

However, this extended context poses a severe efficiency bottleneck. As the trajectory grows, the accumulation of verbose logs, redundant file contents, and obsolete error stacks leads to rapidly growing token consumption. Such overhead results in increased latency, substantial API expenditures, and potentially surpassing the maximum context length of LLMs, which poses a significant barrier to industrial scalability. Furthermore, the ``Lost-in-the-Middle'' phenomenon suggests that feeding excessive noise to the model can degrade its reasoning performance~\cite{LostintheMiddle,RULERLongContextLanguageModels}.

To mitigate this, context compression has become essential. Current research broadly falls into three categories: heuristic-based pruning, which removes historical messages based on rules (e.g., \obsmask{}~\cite{ObsMask}); summarization-based methods, which utilize LLMs to condense history into shorter text (e.g., \agentdiet{}~\cite{AgentDiet}); and selection-based methods, which employ a small model to predict and filter out less important tokens (e.g., \lingua{}~\cite{LLMLingua}). While these methods alleviate the context burden to some extent, they suffer from two critical limitations when applied to the dynamic nature of SE tasks:

\textbf{First, inability to handle dynamic context dependencies.} Existing methods typically employ a ``static, one-pass'' strategy, where compression is performed once based on the agent’s current view of the task, producing a finalized reduced context. For instance, \obsmask{} mechanically removes observations older than a fixed window, while \agentdiet{} performs step-wise trajectory rewriting: at each step, it invokes a cost-efficient LLM to rewrite a previous step (typically a fixed lag behind the current step) into a shorter form. Such approaches are irrevocable and neglect the non-linear nature of debugging. In SE tasks, agents often exhibit focus shifting~\cite{agentdiagnose,Focus-shiftingPatternsofDeveloper}, such as abandoning a hypothesis about a database error to investigate a network configuration mentioned twenty turns earlier. Static methods sever these long-range semantic dependencies because once a block of context is deemed ``irrelevant'' and removed, it cannot be recovered when the agent's focus shifts, leading to context-induced hallucinations.

\textbf{Second, granularity mismatch and precision loss.} There is a trade-off between semantic integrity and compression rate that prior works fail to balance. On one hand, coarse-grained methods like \obsmask{} operate at the message level, forcing the retention of entire verbose logs even if only a single error line is relevant. Conversely, summarization-based methods like \agentdiet{} employ an additional LLM to rewrite context. While this reduces length, the generative nature of summarization poses a severe risk to SE tasks which demand verbatim accuracy. Such methods often abstract away critical details (e.g., specific line numbers or variable names) and, more critically, are prone to hallucinating~\cite{HallucinationDetectioninCodeSummaries} non-existent code behaviors or incorrect error references, thereby misleading the agent. Furthermore, relying on an extra strong LLM for every step incurs prohibitive computational overhead. On the other hand, fine-grained token-level selection (e.g., \lingua{}~\cite{LLMLingua-2}) disregards the syntactic structure of code. Arbitrarily dropping tokens can break JSON objects or function definitions, rendering the context incomprehensible for the LLM.

To address these challenges, we introduce \attncompress{}, a dynamic attention-guided trajectory compression framework designed specifically for the dynamic workflows of SE agents. Our approach bridges the gap between semantic integrity and dynamic adaptability through three key mechanisms:
\begin{enumerate}
    \item \textbf{Structure-Aware Segmentation via PPL Spikes:} To resolve the \textit{granularity mismatch}, we utilize a small proxy model to monitor the Perplexity (PPL) fluctuations~\cite{LongCodeZip} of the output stream. By identifying PPL spikes, which naturally occur at semantic boundaries (e.g., the switch from code to error logs), we segment the trajectory into semantically coherent \textit{blocks} rather than arbitrary tokens. This strategy mitigates the risk of breaking syntactic structures like JSON objects and function bodies after compression.
    
    \item \textbf{Relevance Estimation via Proxy Attention:} Instead of relying on heuristic rules or heavy summarization models, we leverage the attention distribution of a cost-effective small model (e.g., \textit{Qwen3-4B-Instruct} ~\cite{qwen3_4b_instruct_2507_modelcard}). We calculate the attention weights projected from \del{the agent's \textit{current thought}}\note{BQ1} \add{the generation start token of the next response} to historical blocks, enabling us to quantify the precise relevance of each block to the immediate reasoning step.
    
    \item \textbf{Dynamic Context Maintenance:} To overcome the \textit{static} limitation, we introduce a \textit{rolling window} mechanism. Unlike static pruning, our framework maintains a dynamic buffer where recent history is continuously \textit{re-evaluated} against the agent's shifting intent. This allows previously suppressed information to be recalled if it becomes relevant again, effectively mitigating the risk of information loss during goal shifting.
\end{enumerate}

Our contributions are as follows:
\begin{itemize}

    \item We propose a dynamic attention-guided trajectory compression framework that leverages the attention distribution of a cost-effective proxy model to quantify the relevance between historical context and the agent's current work. 
    \item We introduce a PPL-based block segmentation algorithm that preserves the syntactic structure of code and logs during compression.
    \item Extensive evaluation on \swev{}~\cite{SWE-bench} and \mswef{}~\cite{Multi-SWE-bench} demonstrates that \attncompress{} achieves a pass rate of 53.17\%, outperforming prior SOTA \texttt{Agent} \texttt{Diet} while reducing token consumption by 21.6\% and total costs by 33.6\%.

\end{itemize}

\section{Background \& Related Work}
\label{sec:background}

\subsection{LLM-based SE Agents \& Workflow}
\begin{figure}[ht]
    \centering
    \includegraphics[width=0.8\linewidth]{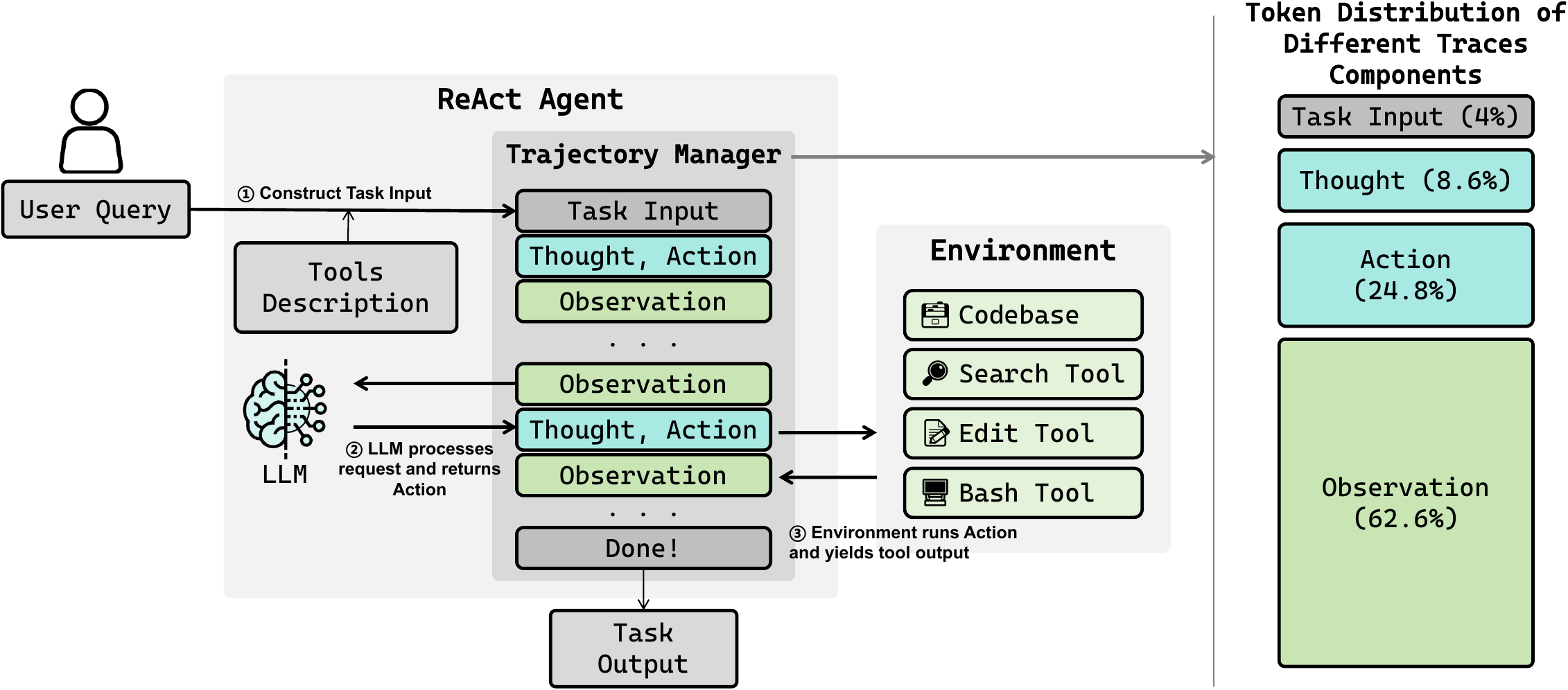}
    \caption{The typical workflow of a ReAct-based SE agent (left) and the token distribution across trace components (right). The statistical analysis reveals that \textit{Observation} content (e.g., logs and code retrieval) dominates the context window (62.6\%), posing a major challenge for long-horizon tasks.}

    \label{fig:workflow}
\end{figure}

In Autonomous Software Engineering (ASE), LLM-based agents are commonly deployed to address GitHub issues, implement feature requests, comprehend codebases, and generate tests in an end-to-end manner~\cite{LLMBasedAgentsforSE_Survey, SWE-agent, SWE-Compass}. As illustrated in the left panel of Figure~\ref{fig:workflow}, the operational workflow typically follows an iterative ReAct (Reasoning and Acting) pattern~\cite{ReAct}. The process begins with 1) constructing task input, where the user query is combined with detailed descriptions of available tools (e.g., search, edit, and bash utilities). Subsequently, the agent enters a cyclic interaction loop:
\begin{itemize}
    \item \textbf{LLM Processing:} Based on the current history, the LLM generates a \textit{Thought} for reasoning and an \textit{Action} to interact with the environment.
    \item \textbf{Environment Execution:} The environment executes the action using specific tools (e.g., searching the codebase or running a script) and yields an \textit{Observation}, which contains the execution results such as file contents or error logs.
\end{itemize}
This cycle repeats until the issue is resolved (``Done!''). While this mechanism enables complex problem-solving, it incurs a significant context overhead. As shown in the statistical breakdown in the right panel of Figure~\ref{fig:workflow}, (derived from the trajectories in RQ1), the distribution of tokens in interaction traces is highly imbalanced. \textbf{Observation} tokens account for the vast majority (62.6\%) of the total context, significantly outweighing \textbf{Action} (24.8\%), \textbf{Thought} (8.6\%), and \textbf{Task Input} (4\%). This data underscores that verbose environmental feedback is the primary contributor to context bloat, making the compression of observations the critical bottleneck for efficiency.

\subsection{The Context Challenge in SE}
\label{sec:2.2}
While extensive context provides necessary information, it introduces two fundamental challenges specific to software engineering tasks: \textit{Information Density Variance} and \textit{Dynamic Relevance}.

\begin{figure}[b]
    \centering
    \begin{subfigure}[c]{0.49\linewidth}
        \centering
        \includegraphics[width=\linewidth]{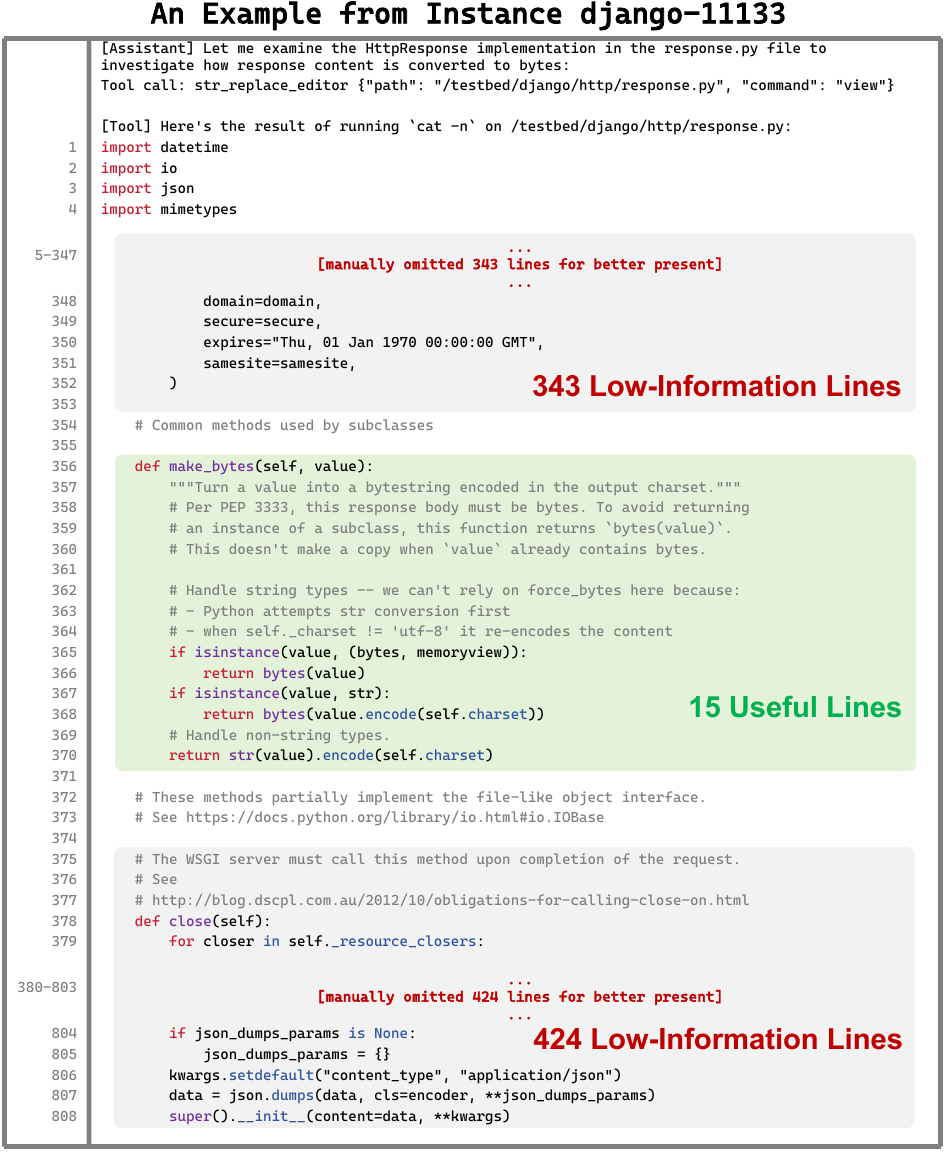}
        \caption{Granularity Mismatch}
        \label{fig:challenge_left}
    \end{subfigure}
    \hfill 
    \begin{subfigure}[c]{0.49\linewidth}
        \centering
        \includegraphics[width=\linewidth]{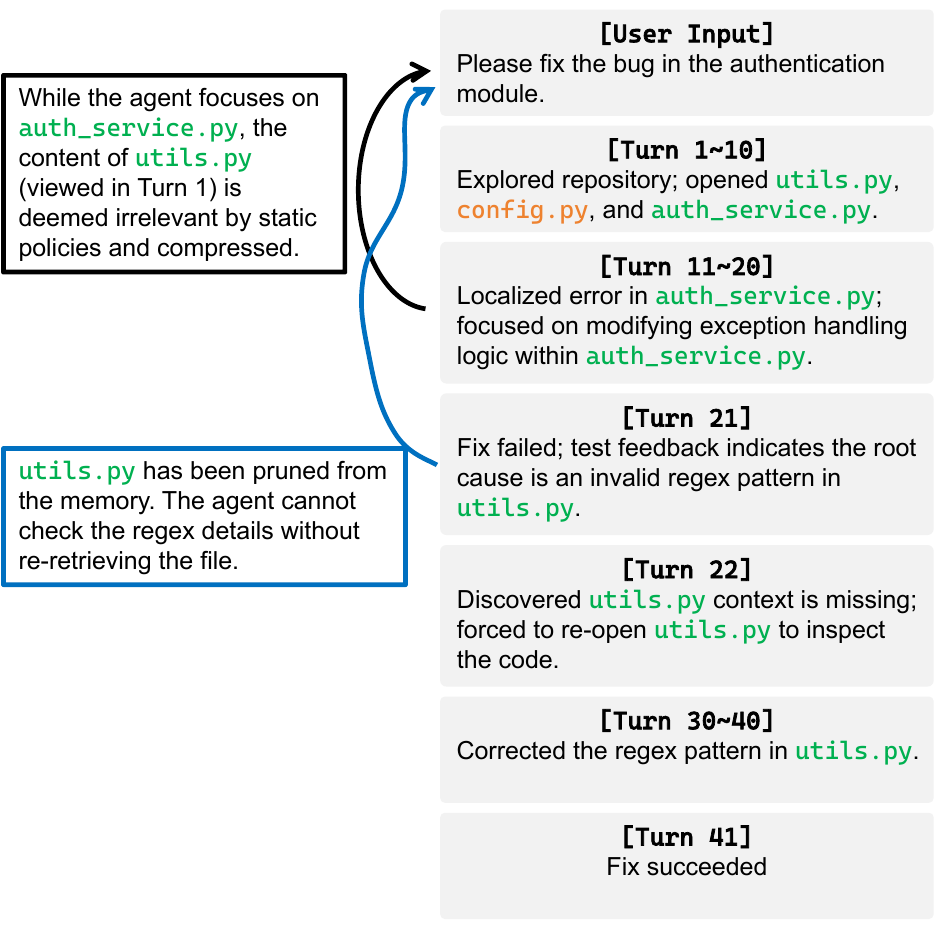}
        \caption{Focus Shifting Problem}
        \label{fig:challenge_right}
    \end{subfigure}
    
    \caption{Challenges in SE context compression. \textbf{(a) Granularity Mismatch:} A case from \texttt{django-11133} showing that within a large file output, only a small function (\texttt{make\_bytes}) is useful. \textbf{(b) Focus Shifting Problem:} A debugging timeline where the agent shifts focus from \texttt{auth\_service.py} to \texttt{utils.py}.}

    \label{fig:challenges}
\end{figure}

\paragraph{Information Density Variance.}
The information density within an agent's trajectory is highly uneven. A significant portion of the context often consists of verbose logs or irrelevant code snippets. 
As illustrated in Figure~\ref{fig:challenge_left}, we present a real-world example from the \swebench{} instance \texttt{django-11133}. The agent retrieves the content of \texttt{response.py} to inspect the \texttt{HttpResponse} implementation. This action generates a massive output where over 700 lines are essentially noise (highlighted in red as ``Low-Information Lines''). The agent effectively needs only the \texttt{make\_bytes} method (approximately 15 lines, highlighted in green) to proceed. 
This variance creates a practical granularity challenge for compression. If compression operates at a coarse unit such as a full message or a full tool output, the method tends to either keep large chunks of noise or discard large chunks that may still contain small but critical details. If compression operates at an overly fine unit such as individual tokens, it risks breaking code and structured logs. This motivates a segmentation strategy that identifies semantically coherent units within long observations so that the agent can retain complete and meaningful parts while discarding irrelevant parts.

\paragraph{Dynamic Relevance.}
The relevance of historical information in SE tasks fluctuates as the agent's debugging focus and hypotheses evolve. As illustrated in Figure~\ref{fig:challenge_right}, the agent initially inspects the whole repository (Turns 1-10), then shifts its focus to \texttt{auth\_service.py} (Turns 11-20) to address the error, believing the root cause lies therein. During this phase, static compression methods (e.g., \obsmask{}, \summary{}, and \agentdiet{}) deem the previously viewed \texttt{utils.py} as irrelevant ``old'' context and permanently prune it. However, in Turn 21, execution feedback reveals that the root cause is actually an incorrect regex pattern defined in \texttt{utils.py}. Since the specific details of \texttt{utils.py} were discarded, the agent cannot ``look back'' to verify the pattern, forcing a redundant re-opening of the file (Turn 22). This inefficiency highlights the need for a dynamic mechanism that can recall previously suppressed blocks when the agent's focus shifts back to them.
\note{CQ5} \add{To quantify how frequently this phenomenon occurs in practice, we randomly sampled 100 trajectories from \summary{} runs and audited them using an LLM-assisted review followed by human verification. We found that 40 of these trajectories exhibited the focus shifting problem, indicating that this challenge is not merely illustrative but occurs in a substantial portion of real agent debugging workflows.}

\subsection{Related Work}
Prior work on context reduction for LLMs and agents can be grouped into three categories. Each category addresses part of the long-context problem but exhibits limitations for SE agent trajectories.

\subsubsection{Selection-based Compression}
Selection-based compression aims to improve LLM efficiency by identifying and retaining only the most informative parts of the input. A significant body of work relies on training dedicated modules to score content relevance. For instance, \texttt{RECOMP}~\cite{RECOMP} trains an abstractive compressor to paraphrase documents, while \texttt{CPC}~\cite{CPC} employs a trained context-aware sentence encoder to judge similarity. Similarly, methods like \texttt{Provence}~\cite{Provence} and \texttt{LLMLingua-2}~\cite{LLMLingua-2} utilize distillation techniques to train classifier models that predict the necessity of individual tokens. Alternatively, non-training approaches such as \texttt{FilCo}~\cite{FilCo} and \texttt{Selective Context}~\cite{SelectiveContext} rely on information-theoretic metrics, calculating the self-information of tokens to prune those with low density. 
However, these approaches encounter two fundamental limitations in the context of SE agents. First, the heavy reliance on specific training restricts their plug-and-play capability and generalizability across diverse SE scenarios. Second, most selection methods operate at a token granularity. Token-level selectors disregard syntactic boundaries, where removing ostensibly low-information tokens, such as brackets, colons, or indentation, which can corrupt the Abstract Syntax Tree (AST) of code or the structure of JSON logs, rendering the context incomprehensible to the LLM. These shortcomings underscore the necessity for a structure-aware compression strategy that respects the semantic coherence of software artifacts.

\subsubsection{Heuristic-based Compression}
Another category relies on simple heuristics, such as keeping only the most recent turns, applying FIFO-style eviction, or masking older observations. \obsmask{}~\cite{ObsMask} is a representative approach that directly drops tool outputs from older dialogue history. Moreover,~\citet{Pan_et_al} propose reformatting code to remove tokens related to whitespaces and indentations. These methods are attractive due to simplicity and low overhead, and they can reduce context length substantially in practice.

However, heuristic strategies typically lack semantic awareness. They assume that old content is less useful, and they do not distinguish between a verbose but unimportant log and a short but critical clue. Meanwhile, as discussed in Section~\ref{sec:2.2}, SE tasks often require revisiting earlier evidence after a hypothesis shift. Heuristic deletion can therefore discard essential information and cause irreversible information loss, reducing agent success rates on complex tasks.

\subsubsection{Summarization-based Compression}
A third family of methods compresses context by utilizing LLMs to summarize or rewrite historical information. 
This strategy is widely adopted in practical agent systems, often in an ad-hoc manner to handle context saturation. For instance, tools like \texttt{Gemini-Cli}~\cite{gemini_cli} and \texttt{Claude Code}~\cite{claude_code} trigger LLM-based compression only when the context window reaches a predefined threshold. Others rely on rigid heuristics: \texttt{Trae Agent}~\cite{traeagent} truncates tool responses to a fixed size (e.g., 16KB), while \texttt{SWE-agent}~\cite{SWE-agent} employs configurable regex patterns to remove specific text blocks. 
In the research domain, \agentdiet{}~\cite{AgentDiet} proposes a more systematic approach. It utilizes an additional LLM to refine or shorten messages, aiming to remove irrelevant parts while keeping key content.

These approaches can preserve high-level semantics, but they introduce new issues in SE settings. First, they strongly depend on the summarization model's ability to retain precise details that are often indispensable for SE tasks, including exact file paths, line numbers, error codes, and variable names. If these details are dropped or altered by hallucinating, the agent may lose grounding and propose incorrect patches. Second, using an extra LLM for summarization increases runtime overhead and latency, and it can complicate deployment when the agent is expected to run at scale or under tight response constraints. 

In summary, existing techniques either prune too aggressively without adapting to dependency shifts, or compress by rewriting content in ways that may lose critical debugging details. These limitations motivate methods that can both preserve structural integrity and dynamically select context based on the agent's current needs. \note{BQ6,CQ3} \add{Although chunking, relevance scoring, and rolling context maintenance are not new in isolation, \attncompress{} differs by unifying PPL-based block segmentation, proxy-attention selection, and dynamic re-evaluation in one training-free middleware built for multi-turn SE agent trajectories, explicitly targeting the granularity mismatch and focus-shifting challenges discussed above.}

\section{Approach}
\label{sec:approach}

\subsection{Overview}
\begin{figure}[t]
    \centering
    \includegraphics[width=0.85\linewidth]{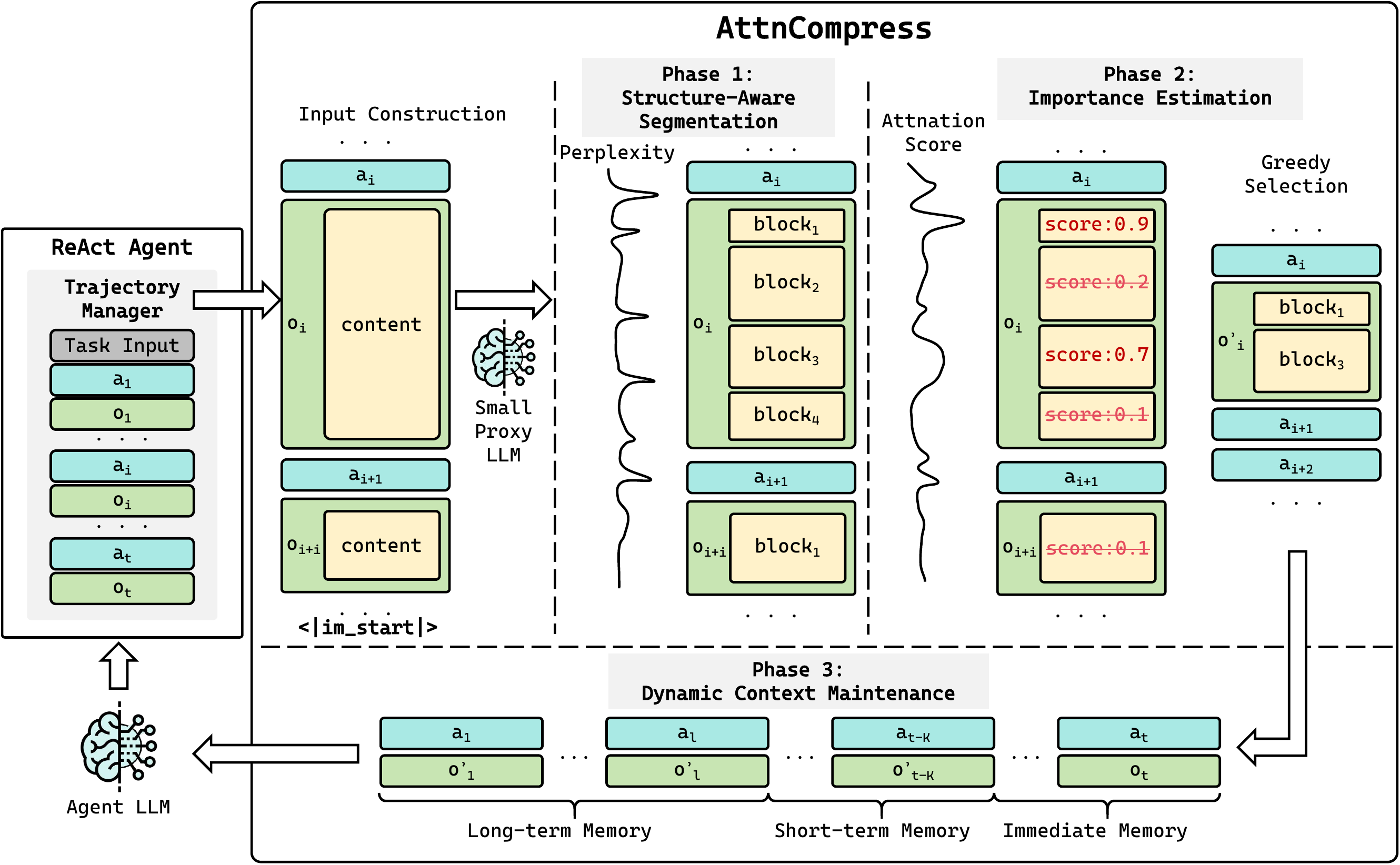}
    \caption{Overview of the \attncompress{} Framework. The system intercepts the raw trajectory, segments observations into blocks, scores them via a proxy model's attention, and dynamically maintains a compressed context using a rolling window mechanism.}
    \label{fig:approach_overview}
\end{figure}

We formalize \attncompress{} as a plug-and-play middleware positioned between the SE Agent's trajectory manager and the backend LLM. As illustrated in Figure~\ref{fig:approach_overview}, the workflow consists of three consecutive phases: \textit{Structure-Aware Segmentation}, \textit{Attention-based Scoring}, \textit{and Dynamic Context Maintenance}.

Given a raw trajectory consisting of multiple interaction turns $T = \{ (a_1, o_1), \dots, (a_t, o_t) \}$, where $a_t$ represents the agent's action and $o_t$ represents the observation from the environment (typically the output of tool executions, often containing verbose code or logs), our goal is to maintain a compressed trajectory $T' = \{ (a_1, o'_1), \dots, (a_t, o'_t) \}$ in real-time. The objective is to ensure $|T'| \ll |T|$ while maximizing the retention of semantic information relevant to the current reasoning step $s_t$, thereby accommodating a longer interaction history within a limited context window.

\subsection{Phase 1: Structure-Aware Segmentation via PPL Spikes}
To address the ``granularity mismatch'' problem in compression, where token-level pruning destroys syntax and message-level pruning retains excessive noise, we introduced an adaptive segmentation algorithm based on Perplexity (PPL) spike detection~\cite{LongCodeZip}. This algorithm utilizes a small model (Proxy Model) as a probe to perceive semantic boundaries within text. \note{MR1, AQ1} \add{Prior work has empirically shown that perplexity-based boundary detection can preserve the syntactic integrity of long code contexts more effectively than fixed-size or heuristic chunking~\cite{LongCodeZip}.}

\paragraph{PPL Calculation.}
For any given tool output text $O$ (e.g., file content read by \texttt{cat} or error traces from \texttt{pytest}), we first split it into lines $L = \{l_1, \dots, l_n\}$. We use the proxy model to calculate the log probability of each token $x$ and compute the average perplexity $PPL(l_i)$ for each line $l_i$:
\begin{equation}
    PPL(l_i) = \exp \left( -\frac{1}{|l_i|} \sum_{x \in l_i} \log P(x \mid x_{<context}) \right)
\end{equation}
Typically, at boundaries where semantic content shifts drastically (e.g., from the end of a function definition to the start of a new one, or from a normal log stream to an error stack), the model exhibits higher ``surprise,'' manifested as local peaks in the PPL curve.

\paragraph{Boundary Detection.}
We determine split boundaries by detecting ``spikes'' in the $PPL$ sequence. To adapt to the fluctuating PPL baselines of different text contents, we employ an adaptive thresholding strategy:
\begin{enumerate}
    \item \textbf{Diff Calculation:} Compute the PPL difference between adjacent lines $D_i = |PPL(l_i) - PPL(l_{i-1})|$.
    \item \textbf{Adaptive Threshold:} Calculate the mean $\mu$ and standard deviation $\sigma$ of the difference sequence. Define the spike threshold $\tau = \mu + h \cdot \sigma$, where $h$ is a sensitivity coefficient.
    \item \textbf{Segmentation:} Mark all local maxima points where $D_i > \tau$ as split boundaries. Additionally, to avoid fragmentation, minute blocks with length only one line are merged with the preceding block.
\end{enumerate}
Finally, the raw output $O$ is transformed into a series of semantic blocks $B = \{b_1, b_2, \dots, b_m\}$, where each block acts as a relatively independent unit in terms of syntax or semantics.
\note{MR1, AQ4} \add{Besides, introducing overlapping regions between adjacent blocks is a feasible engineering improvement that could further improve cross-boundary coherence, but it would introduce additional hyperparameters (e.g., overlap size). We therefore adopt only the simplest non-overlapping segmentation scheme in this paper.}

\subsection{Phase 2: Importance Estimation via Proxy Attention}
\note{MR1, AQ1} \add{To evaluate the dynamic relevance of historical blocks to the current task, we leverage the attention weights of the proxy model as a filtering signal, which prior work has shown can identify tokens semantically salient to a query~\cite{AttnComp}.}

\paragraph{Input Construction.}
To calculate relevance, we construct an input sequence that incorporates the current context. The input sequence is formed by concatenating three components: (1) Context History (using the compressed trajectory $T'_{1:t-1}$ or raw trajectory $T_{1:t-1}$), (2) The New Observation $O_{new}$, and (3) a single special Query Token $q_{gen}$ appended at the very end (e.g., \del{a conversation start token like \texttt{<|im\_start|>} representing the beginning of the agent's next thought}\note{MR1, BQ1} \add{the generation start token in the chat template, such as \texttt{<|im\_start|>}, which corresponds to $T[t].start\_token$ in Algorithm~\ref{alg:context_maintenance}}). \add{In a causal LLM, this token must attend to the preceding context before generating the agent's next response.} This construction allows the model to aggregate the attention from the entire preceding context onto this final token, representing the agent's ``current reasoning state."
\paragraph{Scoring Metric.}
We feed the constructed sequence into the proxy model to perform a forward pass. Crucially, this process is computationally efficient: the logits required for PPL-based segmentation (Phase 1) and the attention maps required for scoring (Phase 2) are extracted simultaneously in a single inference pass. We utilize the attention map from a specific layer (the last layer is used in our experiments). For each candidate block $b_i$, its importance score $Score(b_i)$ is defined as the average attention weight projected from the single query token $q_{gen}$ to the tokens within the block:
\begin{equation}
Score(b_i) = \frac{1}{|b_i|} \sum_{t \in b_i} A(q_{gen}, t)
\end{equation}
Here, $A(q_{gen}, t)$ denotes the attention weight from the query token $q_{gen}$ to a token $t$ inside block $b_i$. This metric effectively captures the semantic alignment between the agent's current state and the historical information.

\paragraph{Selection Strategy.}
Based on the calculated scores, we apply a \textit{Greedy Selection} strategy. Let $B=\{b_1,\dots,b_m\}$ be all candidate blocks segmented from the tool output context, and let $|b_i|$ denote the number of tokens in block $b_i$. Given a compression ratio $\rho\in(0,1)$, we set a token budget $L_{\text{budget}} = \rho \cdot \sum_{i=1}^{m} |b_i|$. 
We then sort all blocks in descending order of $Score(b_i)$ and greedily add blocks to the retention set until the accumulated retained tokens reach $L_{\text{budget}}$. Unselected blocks are discarded.

\subsection{Phase 3: Dynamic Rolling Maintenance}
In SE tasks, the agent's focus shifts continuously as the programming process evolves, leading to \textit{task focus drift}. Traditional static compression (compress once, delete forever) results in the inability to retrieve old information. To address this, we design a Three-Tier Rolling Window mechanism to maintain context dynamically.

\paragraph{The Context Buffer.}
As shown in Figure~\ref{fig:approach_overview}, at interaction turn $t$, we partition the trajectory into three regions:
\begin{equation}
    T = \{ \underbrace{(a_1, o_1), \dots, (a_l, o_l)}_{\text{Long-term}}, \underbrace{\dots, (a_{t-k}, o_{t-k})}_{\text{Short-term}}, \underbrace{\dots, (a_t, o_t)}_{\text{Immediate}} \}
\end{equation}

\begin{enumerate}
    \item \textbf{Immediate Memory (Raw):} The most recent $k$ turns (e.g., Tail 2) are kept in their raw state without compression. This ensures the agent maintains coherent perception of the immediate interaction, preventing the loss of actionable details (e.g., filenames, line numbers).
    \item \textbf{Short-term Memory (Rolling Buffer):} The intermediate region covering turns $I_{long}+1$ through $t-k$. This serves as a dynamic buffer. In every interaction turn, all content within this region is re-evaluated against the current $Q_{curr}$, and attention scores are re-calculated to re-compress the content. This ensures the agent can extract the most relevant information from recent history based on its latest intent.
    \item \textbf{Long-term Memory (Fixed Archive):} The region from turn $1$ to $I_{long}$. This serves as the archived history. To minimize computational overhead, this region remains static during standard interaction steps, which allows the LLM to leverage prefix caching~\cite{vllm}. Consequently, we avoid reconstructing this long-term memory at every step, significantly reducing the inference cost. This archive is only reconstructed when a ``Global Refresh'' is triggered.
\end{enumerate}


\paragraph{Periodic Global Refresh.}
To balance computational cost with recall capability, we introduce a sliding-window-based Global Refresh mechanism (Algorithm~\ref{alg:context_maintenance}). This mechanism operates in two distinct modes:

\textbf{Mode 1: Incremental Update (Standard).}
This corresponds to the else block (Lines 19-25). In most interaction steps where the buffer is not full, the long-term boundary $I_{long}$ remains unchanged.
We perform selection and re-compression only on the short-term region $T_{short}$ (Lines 19-22), generating $T'_{short}$. Crucially, as shown in Line 25, the final context is constructed by concatenating the pre-existing compressed archive $T'_{long}$ with the newly updated short-term blocks. Since $T'_{long}$ is textually invariant, this strategy maximizes the cache hit rate for the Agent LLM's prefix cache.

\textbf{Mode 2: Global Refresh (Triggered).}
When the short-term buffer accumulates beyond threshold $M$ (condition at Line 9), a Global Refresh is triggered (Lines 10-17).
To address the limitation of a frozen archive, we merge the existing long-term archive with the current short-term buffer ($T_{long} + T_{short}$ in Line 11) and perform a global selection over this combined history. This allows the agent to ``resurrect'' previously discarded blocks from the deep history if they become relevant to the current task.
Finally, we advance the archive pointer $I_{long}$ to the current raw boundary $I_{raw}$ (Line 16), effectively committing the current history to the long-term archive. While this step invalidates the prefix cache, it is performed infrequently to minimize overhead while ensuring semantic completeness.

This mechanism effectively combines high-frequency updates for the short term (adapting to rapid focus changes) with low-frequency refreshes for the long term (adapting to major shifts in debugging direction), solving the focus drift problem while maintaining efficiency.

\begin{algorithm}[h]
\caption{Dynamic Context Maintenance with Rolling Window}
\label{alg:context_maintenance}
\resizebox{0.95\linewidth}{!}{ 
\begin{minipage}{\linewidth} 
\begin{algorithmic}[1]
\Require Full Trajectory $T$, Compressed Trajectory $T'$, Current Turn $t$, Tail Size $k$, Rolling Window Size $M$
\State $I_{long} \gets \text{Index of the last long memory turn}$ \Comment{Initially 0}
\State $I_{raw} \gets t - k$

\State \textbf{// Phase 3.1: Define Context Regions}
\State $T_{long} \gets T[0 : I_{long}]$
\State $T'_{long} \gets T'[0 : I_{long}]$
\State $T_{short} \gets T[I_{long} : I_{raw}]$
\State $T_{raw} \gets T[I_{raw} : t]$ \Comment{Keep Raw}

\State \textbf{// Phase 3.2: Check Trigger for Global Refresh}
\If{Length($T_{short}$) $\ge M$}
    \State \textbf{// Global Refresh: Re-evaluate EVERYTHING before Raw}
    \State $Candidates \gets \text{Segment}(T_{long} + T_{short})$
    \State $Scores \gets \text{ProxyAttention}(T_{long},T_{short},T_{raw}, Query=T[t].start\_token)$
    \State $Selected \gets \text{SelectTop}(Candidates, Scores)$
    \State $T'_{compressed} \gets \text{Reconstruct}(Selected)$
    
    \State \textbf{// Update Archive Pointer}
    \State $I_{long} \gets I_{raw}$
    \State \textbf{Output Context:} $T'_{compressed} + T_{raw}$
\Else
    \State \textbf{// Local Rolling: Re-evaluate only Short-term}
    \State $Candidates \gets \text{Segment}(T_{short})$
    \State $Scores \gets \text{ProxyAttention}(T'_{long},T_{short},T_{raw}, Query=T[t].start\_token)$
    \State $Selected \gets \text{SelectTop}(Candidates, Scores)$
    \State $T'_{short} \gets \text{Reconstruct}(Selected)$
    
    \State \textbf{// Merge with existing Long-term}
    \State \textbf{Output Context:} $T'_{long} + T'_{short} + T_{raw}$
\EndIf
\end{algorithmic}
\end{minipage}
} 
\end{algorithm}

\subsection{Implementation}
\attncompress{} is a generalized trajectory compression framework designed to be compatible with various ReAct-style LLM agents. To evaluate its effectiveness in a SOTA setting, we integrated \attncompress{} into \texttt{Trae-Agent}~\cite{traeagent}, a leading open-source agent that has demonstrated top-tier performance on benchmarks like \swebench{} Verified~\cite{swebenchverified, SWEbenchLeaderboards}.

For the proxy model ($LLM_{proxy}$), we selected \textit{Qwen3-4B-Instruct}~\cite{Qwen3, qwen3_4b_instruct_2507_modelcard}. We chose this model for two strategic reasons: first, despite its small size, it retains strong capabilities in code syntax understanding and context processing; second, its significantly lower parameter count incurs minimal computational overhead, satisfying the real-time requirements of the agent workflow. All experiments, including the inference of the proxy model and the execution of the agent loop, were conducted on a server equipped with 8 NVIDIA A100 GPUs. 

\attncompress{} involves several key hyperparameters. To determine the optimal configuration, we conducted a preliminary ablation study on a subset of the \swev{} dataset (randomly select 100 instances). Based on the trade-offs between token cost and pass rate, we established the following default settings for our main evaluation:
\begin{itemize}
    \item \textbf{Tail Size ($k$):} 2 (The 2 most recent turns are always kept raw).
    \item \textbf{Compression Ratio ($\rho$):} 0.2 (Retaining top 20\% of information by attention score).
    \item \textbf{PPL Block Threshold:} -2 (Used for adaptive boundary detection).
    \item \textbf{Attention Layer:} -1 (Using the attention map from the last layer of the proxy model).
    \item \textbf{Rolling Window Size ($M$):} 10 (Triggers a Global Refresh when the short-term buffer accumulates 10 turns).
\end{itemize}
We omit the detailed analysis of how these parameters affect performance in this section. A comprehensive sensitivity analysis and the impact of different hyperparameter combinations are presented in Section~\ref{sec:rq2}.

\section{Experimental Design}

\subsection{Research Questions}
To systematically assess the proposed framework, we structure our evaluation around three key research questions:

\paragraph{RQ1: Cost-Effectiveness Trade-off.} 
\textbf{Does \attncompress{} achieve a superior balance between problem-solving effectiveness and computational efficiency compared to state-of-the-art baselines?}
We investigate whether \attncompress{} can maintain or improve the pass rate on \swev{} while reducing token consumption, monetary cost, and end-to-end latency compared to existing methods.

\paragraph{RQ2: Component Contribution.} 
\textbf{How do the internal architectural components and hyperparameters impact the performance of \attncompress{}?}
We perform an ablation study to quantify the individual contributions of the ppl-based segmentation, proxy-attention scoring, and rolling window mechanism, while also analyzing the sensitivity of key hyperparameters (e.g., tail size $k$ and compression ratio $\rho$).

\paragraph{RQ3: Generalization Capabilities.} 
\textbf{Does the framework generalize well across different proxy models and datasets?}
We assess the robustness of \attncompress{} by evaluating its performance with different proxy model (e.g., \textit{Qwen} vs. \textit{Llama}) and testing its adaptability on the multilingual \mswebench{} dataset.

\subsection{Experimental Setup}

\subsubsection{Datasets}
\label{sec:datasets}
We evaluate \attncompress{} and the baselines on two complementary benchmarks to assess both effectiveness and generalization, consistent with the setup in~\cite{AgentDiet}.

\begin{itemize}
    \item \textbf{\swev{}~\cite{swebenchverified}:} This is the primary dataset for our evaluation. It consists of 500 human-verified software engineering tasks derived from real-world GitHub issues. To maintain consistency and eliminate sampling bias, we do not sample a new subset; instead, we utilize the exact instance lists curated by \agentdiet{}~\cite{AgentDiet}. Specifically, we use their defined validation set (100 instances) for the parameter tuning and ablation studies described in RQ2. The remaining test set (200 instances), as identified in their work, is strictly reserved for the main comparative evaluation in RQ1 and the proxy-model generalization study in RQ3.
    
    \item \textbf{\mswef{}~\cite{Multi-SWE-bench, multiswebenchflash_dataset}:} To evaluate the generalization capabilities of our approach across different languages and environments (RQ3), we utilize \mswef{}. This benchmark contains 300 instances covering seven programming languages (Rust, TypeScript, JavaScript, Java, Go, C, and C++). These tasks typically present higher complexity, often requiring the agent to troubleshoot environment build errors, providing a robust testbed for the agent's adaptability.
\end{itemize}

\subsubsection{Baselines}
We compare \attncompress{} against seven baselines. All baselines are integrated into the same Trae-Agent~\cite{traeagent} framework to ensure a fair comparison.

\begin{itemize}
    \item \textbf{\original{}:} The unmodified Trae-Agent that retains the full interaction history. This serves as the upper bound for context completeness and the lower bound for efficiency.

    \item \textbf{\random{}:} A sanity check that randomly drops 75\% of the tokens from previous turns.
    \item \note{MR1, AQ6} \add{\textbf{\sliding{}:} A strict token-budget sliding window baseline that retains only the most recent trajectory content under a fixed token budget. To keep its token cost comparable to the other compression methods, we set the window size to 8k tokens.}
    \item \textbf{\obsmask{}~\cite{ObsMask}:} A rule-based approach that masks the outputs of tools from older turns with placeholder, assuming that recent observations are more relevant.

    \item \textbf{\lingua{}~\cite{LLMLingua-2}:} A token compression method that uses a small BERT-based~\cite{BERT} model to classify and remove non-essential tokens.

    \item \textbf{\summary{}~\cite{ACON}:} A standard summarization approach where an additional LLM is prompted to summarize the history of multiple turns into a concise paragraph.
    \item \textbf{\agentdiet{}~\cite{AgentDiet}:} A recent SOTA method that employs a reflection module (i.e., a summarize LLM) to rewrite and condense the trajectory after each step.
\end{itemize}

For all baseline methods, we generally adopt the default hyperparameter settings reported in their original works. However, to ensure a strictly fair comparison, we standardize the tail size parameter (the number of most recent turns retained in raw format) to $k=2$ across all applicable methods. This alignment follows the experimental protocol of \agentdiet{}~\cite{AgentDiet}, ensuring that observed performance differences stem from the compression strategy rather than the amount of immediate raw history. Meanwhile, for the summarization-based methods (\summary{} and \agentdiet{}), we utilize the identical LLM backbone as the agent LLM to perform the summarization and reflection tasks. This ensures that these baselines operate at their optimal capability and are not bottlenecked by a weaker summary model.

\subsubsection{Metrics}
We utilize a comprehensive set of metrics to evaluate the trade-off between performance and cost.

\begin{itemize}
    \item \textbf{\textit{Pass\%}:} The ratio of successfully resolved instances in the benchmark. This is the primary indicator of whether compression harmed the agent's reasoning capabilities.
    \item \textbf{\textit{Step}:} The average number of interaction turns required to solve a task. An increase in steps typically indicates that the agent lost critical information due to compression and had to perform redundant actions to recover it.
    \item \textbf{\textit{PStep}:} The average number of interaction turns required for \textit{only successfully resolved} instances. Unlike \textbf{\textit{Step}}, this metric isolates the efficiency of successful trajectories, filtering out noise from failed attempts that simply exhaust the maximum step limit.
    \item \textbf{\textit{Input (I)} \& \textit{Output (O)}:} The accumulated number of input and output tokens used by the backend LLM. \note{MR2, AQ3, BQ3} \add{We first sum input and output tokens over all interaction turns within each instance, and then report the average of these per-instance totals across instances.}
    \item \textbf{\textit{Agent Cost} ($C_{agent}$):} The monetary cost incurred solely by the \textbf{backend SE agent} during interaction steps. This reflects the direct expenditure on reasoning and generation based on the (potentially compressed) context.
    \item \textbf{\textit{Compression Cost} ($C_{comp}$):} The cost incurred specifically by the compression mechanism itself (e.g., the summarizer cost in \agentdiet{}). \note{MR2, BQ5, CQ4} \add{Note that while the proxy model in \attncompress{} can be deployed locally to avoid API charges, we calculate $C_{comp}$ based on its official API pricing~\cite{qwen3_4b_instruct_2507_modelcard} to ensure a fair comparison.}
    \item \textbf{\textit{Total Cost} ($C_{total}$):} The sum of the agent cost and compression cost ($C_{total} = C_{agent} + C_{comp}$). Crucially, all cost calculations account for the input token discounts provided by prefix caching to reflect realistic API pricing. 

\end{itemize}

\section{Results}
\label{sec:results}

In this section, we present the experimental results to answer the three research questions proposed in Section 4.1. 

\subsection{RQ1: Cost-Effectiveness Trade-off}

To demonstrate the superiority of \attncompress{}, we evaluated its performance on \textit{SWE-bench Verified} using three different backend agents: \textit{Qwen3-Coder-30B}~\cite{qwen3coder30bA3binstruct}, \textit{Qwen3-235B-Instruct}~\cite{qwen3235bA22binstruct2507}, and \textit{Gemini-3-Flash}~\cite{gemini_3_flash_modelcard}. Table~\ref{tab:main_results} summarizes the comprehensive results.

\begin{table*}[ht]
\caption{Main Results on SWE-bench Verified. We report Pass Rate (Pass\%), Input/Output token usage (k), Agent Cost (\$), Compression Overhead (\$), Total Cost (\$), Average Steps, and Steps for Passed instances (PStep). \note{BQ3} \add{All token, cost, and step values are per-instance averages.} \textbf{Bold} indicates the best performance among compression methods; \underline{Underline} indicates the second best.}
\label{tab:main_results}
\resizebox{0.9\textwidth}{!}{%
\begin{tabular}{lccccccccc}
\toprule
\textbf{Method} & \textbf{Agent LLM} & \textbf{\textit{Pass (\%)}} & \textbf{\textit{Input (k)}} & \textbf{\textit{Output (k)}} & \textbf{\boldmath $C_{agent}$} & \textbf{\boldmath $C_{comp}$} & \textbf{\boldmath $C_{total}$} & \textbf{\textit{Step}} & \textbf{\textit{PStep}} \\ \midrule
\multirow{4}{*}{\original{}} 
 & Gemini-3-Flash & 72.50 & 1150.97 & 6.13 & 0.2156 & / & 0.2156 & 50.93 & 47.31 \\
 & Qwen3-235B & 46.50 & 1278.10 & 6.31 & 0.0864 & / & 0.0864 & 45.62 & 35.84 \\
 & Qwen3-Coder-30B & 46.50 & 925.41 & 10.74 & 0.0546 & / & 0.0546 & 41.24 & 34.51 \\ \cmidrule(l){2-10} 
 & \textit{Mean} & 55.17 & 1118.16 & 7.73 & 0.1189 & / & 0.1189 & 45.93 & 39.22 \\ \midrule

\multirow{4}{*}{\random{}} 
 & Gemini-3-Flash & 67.00 & 1075.24 & 6.69 & 0.1896 & / & 0.1896 & 66.97 & 61.58 \\
 & Qwen3-235B & 39.00 & 875.20 & 7.62 & 0.0679 & / & 0.0679 & 60.81 & 43.55 \\
 & Qwen3-Coder-30B & 38.50 & 762.83 & 13.27 & 0.0616 & / & 0.0616 & 56.50 & 44.65 \\ \cmidrule(l){2-10} 
 & \textit{Mean} & 48.17 & 904.42 & 9.19 & 0.1064 & \textbf{/} & 0.1064 & 61.42 & 49.93 \\ \midrule

\multirow{4}{*}{\lingua{}} 
 & Gemini-3-Flash & 69.50 & 906.21 & 6.43 & 0.1469 & / & 0.1469 & 62.84 & 58.03 \\
 & Qwen3-235B & 36.50 & 799.10 & 7.10 & 0.0688 & / & 0.0688 & 55.57 & 39.07 \\
 & Qwen3-Coder-30B & 40.00 & 745.77 & 13.17 & 0.0578 & / & 0.0578 & 56.53 & 42.70 \\ \cmidrule(l){2-10} 
 & \textit{Mean} & 48.67 & 817.03 & 8.90 & 0.0912 & \textbf{/} & \underline{0.0912} & 58.31 & 46.60 \\ \midrule

\multirow{4}{*}{\summary{}} 
 & Gemini-3-Flash & 67.00 & 896.45 & 5.76 & 0.1731 & 0.0185 & 0.1916 & 61.75 & 53.86 \\
 & Qwen3-235B & 42.50 & 836.14 & 7.56 & 0.0801 & 0.0089 & 0.0890 & 51.63 & 34.29 \\
 & Qwen3-Coder-30B & 43.00 & 698.96 & 10.98 & 0.0509 & 0.0063 & 0.0572 & 44.24 & 35.23 \\ \cmidrule(l){2-10} 
 & \textit{Mean} & 50.83 & 810.52 & 8.10 & 0.1014 & 0.0112 & 0.1126 & 52.54 & \underline{41.13} \\ \midrule

\multirow{4}{*}{\obsmask{}} 
 & Gemini-3-Flash & 67.50 & 775.20 & 6.29 & 0.0699 & / & 0.0699 & 71.39 & 66.27 \\
 & Qwen3-235B & 31.50 & 634.67 & 8.82 & 0.0350 & / & 0.0350 & 67.05 & 39.32 \\
 & Qwen3-Coder-30B & 42.50 & 476.91 & 10.38 & 0.0282 & / & 0.0282 & 51.99 & 43.24 \\ \cmidrule(l){2-10} 
& \textit{Mean} & 47.17 & \underline{628.93} & 8.49 & \textbf{0.0443} & \textbf{/} & \textbf{0.0443} & 63.47 & 49.61 \\ \midrule

\multirow{4}{*}{\add{\sliding{}}}
& Gemini-3-Flash & 67.00 & 641.79 & 5.75 & 0.1572 & / & 0.1572 & 57.63 & 51.50 \\
& Qwen3-235B & 42.00 & 602.30 & 6.53 & 0.0766 & / & 0.0766 & 52.12 & 36.20 \\
& Qwen3-Coder-30B & 40.50 & 520.08 & 10.21 & 0.0567 & / & 0.0567 & 43.90 & 36.49 \\ \cmidrule(l){2-10}
& \textit{Mean} & 49.83 & \textbf{588.06} & 7.50 & 0.0968 & \textbf{/} & 0.0968 & \underline{51.21} & 41.40 \\ \midrule

\multirow{4}{*}{\agentdiet{}} 
 & Gemini-3-Flash & 68.00 & 713.40 & 6.31 & 0.1400 & 0.0932 & 0.2332 & 51.93 & 47.15 \\
 & Qwen3-235B & 43.00 & 1143.56 & 6.26 & 0.0838 & 0.0423 & 0.1260 & 49.96 & 35.14 \\
 & Qwen3-Coder-30B & 42.50 & 610.37 & 9.88 & 0.0424 & 0.0269 & 0.0693 & 42.49 & 35.00 \\ \cmidrule(l){2-10} 
 & \textit{Mean} & \underline{51.17} & 822.44 & \underline{7.49} & \underline{0.0887} & 0.0541 & 0.1429 & \textbf{48.12} & \textbf{39.10} \\ \midrule

\multirow{4}{*}{\textbf{\attncompress{}}} 
 & Gemini-3-Flash & 71.50 & 780.17 & 5.45 & 0.1543 & \add{0.0027} & \add{0.1570} & 60.46 & 55.48 \\
 & Qwen3-235B & 43.50 & 619.19 & 6.21 & 0.0706 & \add{0.0020} & \add{0.0726} & 50.53 & 36.00 \\
 & Qwen3-Coder-30B & 44.50 & 534.62 & 10.20 & 0.0532 & \add{0.0019} & \add{0.0551} & 46.30 & 38.56 \\ \cmidrule(l){2-10} 
& \textit{Mean} & \textbf{53.17} & 644.66 & \textbf{7.29} & 0.0927 & \add{0.0022} & \add{0.0949} & 52.43 & 43.35 \\ \bottomrule
\end{tabular}%
}
\end{table*}

\paragraph{Pass Rate Analysis.}
The primary challenge in context compression is balancing information retention with token reduction. As observed in Table~\ref{tab:main_results}, applying any compression strategy naturally results in a slight performance dip compared to the \original{} full-context baseline (Mean Pass 55.17\%). However, relying on full context is often impractical or impossible. Most current LLMs are constrained by context windows, which are easily exceeded by the massive observation logs generated during long-horizon SE tasks. Consequently, compression strategies are necessary not merely for cost savings, but to enable the agent to function within these hard limits.

Among all compression techniques, \attncompress{} achieves the best balance between performance and overhead. It attains the highest mean Pass Rate of \textbf{53.17\%}, outperforming heuristic methods like \obsmask{} (47.17\%) \note{AQ6} \add{and \sliding{} (49.83\%)}, as well as selection-based methods like \lingua{} (48.67\%). \note{BQ4} \add{Moreover, \attncompress{} consistently outperforms all baselines individually across all three agent LLMs (achieving 71.50\%, 43.50\%, and 44.50\% on Gemini-3-Flash, Qwen3-235B, and Qwen3-Coder-30B, respectively).} 
Compared to the previous SOTA summarization method \agentdiet{} (51.17\%), \attncompress{} achieves a \textbf{3.9\%} relative improvement in mean pass rate, demonstrating better information retention.

\note{MR2, BQ2} \add{To further assess statistical robustness, we repeated the \swev{} (200 case subset) evaluation five times using \textit{Qwen3-Coder-30B} as the backend agent. As shown in Table~\ref{tab:multi_runs}, \attncompress{} achieves a higher mean pass rate than both \summary{} (43.70\% vs. 42.70\%) and \agentdiet{} (43.70\% vs. 41.80\%), suggesting that the improvement is not driven by a single lucky run.}

\begin{table}[h]
\caption{\note{MR2, BQ2} \add{Repeated-run robustness on \swev{} with \textit{Qwen3-Coder-30B}. We report pass-rate and total-cost statistics over five runs.}}
\label{tab:multi_runs}
\resizebox{0.75\linewidth}{!}{%
\begin{tabular}{lcccccccc}
\toprule
\multirow{2}{*}{\textbf{Method}} & \multicolumn{4}{c}{\textbf{\textit{Pass (\%)}}} & \multicolumn{4}{c}{\textbf{\boldmath $C_{total}$ (\$)}} \\ \cmidrule(lr){2-5} \cmidrule(l){6-9}
& \textbf{\textit{Mean}} & \textbf{\textit{Std}} & \textbf{\textit{Min}} & \textbf{\textit{Max}} & \textbf{\textit{Mean}} & \textbf{\textit{Std}} & \textbf{\textit{Min}} & \textbf{\textit{Max}} \\ \midrule
\original{} & 44.10 & 2.38 & 41.00 & 46.50 & 0.0563 & 0.0016 & 0.0546 & 0.0582 \\
\summary{} & 42.70 & 2.22 & 40.00 & 46.00 & 0.0585 & \textbf{0.0012} & 0.0572 & 0.0599 \\
\obsmask{} & 40.80 & \textbf{2.08} & 38.00 & 43.00 & \textbf{0.0305} & 0.0013 & \textbf{0.0282} & \textbf{0.0313} \\
\sliding{} & 39.20 & 2.41 & 36.00 & 42.00 & 0.0567 & \textbf{0.0012} & 0.0556 & 0.0584 \\
\agentdiet{} & 41.80 & 2.99 & 38.50 & 45.50 & 0.0718 & 0.0028 & 0.0688 & 0.0751 \\
\textbf{\attncompress{}} & \textbf{43.70} & 2.17 & \textbf{40.50} & \textbf{46.50} & 0.0566 & 0.0013 & 0.0551 & 0.0584 \\ \bottomrule
\end{tabular}%
}
\end{table}

\note{AQ2} \add{To understand the remaining gap from the full-context baseline, we manually inspected the failed cases. The main failure mode is that \attncompress{} can still remove information that later becomes important, such as an earlier file path, helper function, or error message. When the agent's subsequent reasoning depends on such omitted evidence, it may make an incorrect decision and fail to recover within the remaining steps.}

Moreover, \attncompress{} demonstrates superior cost efficiency compared to both the full-context and SOTA approaches. By avoiding the expensive process of using a large LLM to summarize every turn (as in \agentdiet{}), \attncompress{} lowers the total cost to \textbf{0.0949\$}, representing a \textbf{33.6\%} reduction compared to \agentdiet{} (0.1429\$) and a \textbf{20.2\%} reduction compared to the \original{} baseline (0.1189\$). \note{BQ4} \add{This cost efficiency is robust across all models; for example, on the expensive \textit{Qwen3-235B} model, \attncompress{} reduces total cost from 0.0864\$ (\original{}) to 0.0726\$.} Furthermore, regarding token consumption, \attncompress{} operates with a significantly leaner context (Mean Input: 644.66k). This corresponds to a \textbf{21.6\%} reduction against \agentdiet{} (822.44k) and a substantial \textbf{42.3\%} reduction against \original{} (1118.16k), validating that our attention-based selection mechanism efficiently filters noise in raw contexts.


\paragraph{Efficiency and Latency.}

\begin{figure}[ht]
    \centering
    \includegraphics[width=0.5\linewidth]{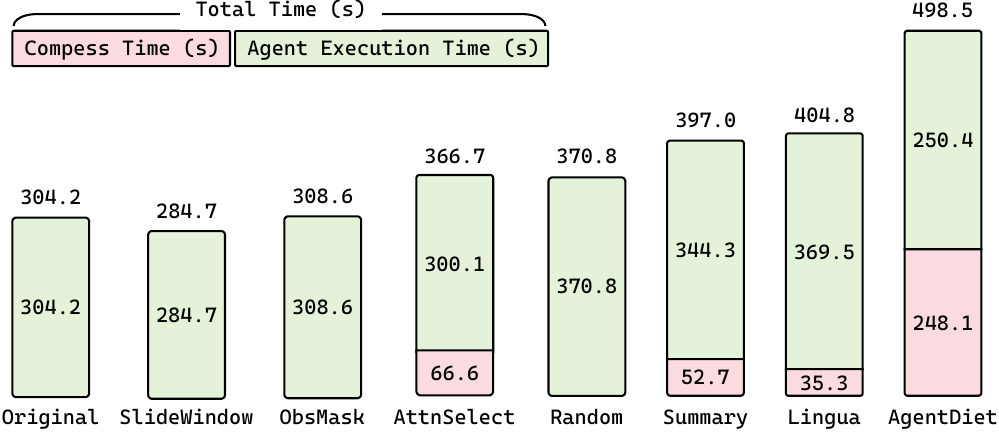} 
    \caption{End-to-end Latency Breakdown.}
    \label{fig:latency_breakdown}
\end{figure}

Beyond monetary cost, latency is a critical factor for real-time agents. We analyzed the time overhead using the breakdown illustrated in Figure~\ref{fig:latency_breakdown}.

\attncompress{} incurs a manageable computational overhead with an average total end-to-end time of \textbf{366.7s}, where the compression (proxy model inference) accounts for only \textbf{66.6s}. This is significantly faster than the SOTA method \agentdiet{}, which requires an average of \textbf{498.5s} due to substantial summarization overhead (248.1s). While heuristic methods like \obsmask{} and \sliding{} are indeed faster (total time 303.6s and 284.7s, respectively) due to zero processing cost, they suffer from a significantly lower pass rate (47.17\% and 49.83\%, respectively). \attncompress{} thus strikes a critical balance, offering a reasonable trade-off between the speed of heuristic pruning and the effectiveness of heavy summarization.

Interestingly, our method also outperforms lighter baselines like \random{} and \lingua{} in terms of total efficiency, despite those methods having negligible compression overhead. As shown in the \textbf{\textit{Step}} column of Table~\ref{tab:main_results}, those two methods often disrupt semantic continuity, causing the agent to get confused. This forces the agent to perform redundant actions to recover missing information, driving up the average step count (\random{}: 61.42 steps; \lingua{}: 58.31 steps). In contrast, \attncompress{} preserves the semantic integrity of the trajectory, allowing the agent to solve problems in fewer steps (52.43 steps), thereby reducing the total cumulative inference time of the backend agent.

\mybox{Conclusion 1}{
\attncompress{} achieves the highest pass rate among studied compression methods (53.17\%), improving upon prior SOTA methods by 3.9\% while reducing total costs by over 33.6\%. Furthermore, it lowers end-to-end latency by 26.4\% compared to \agentdiet{}, offering more practical solution for long-horizon SE tasks.
}

\subsection{RQ2: Component Contribution}
\label{sec:rq2}
To quantify the contribution of individual architectural components to the overall performance, we conducted an ablation study on a subset of \swev{} as detailed in Section~\ref{sec:datasets} with \textit{Qwen3-Coder-30B} as agent LLM. We subsequently analyzed the sensitivity of the framework to key hyperparameters to determine the optimal configuration.

\paragraph{Analysis of PPL-based Segmentation.}
We first evaluate the necessity of our structure-aware segmentation by replacing PPL-based blocking with standard token-level pruning (while maintaining the same compression ratio). As shown in Table~\ref{tab:ablation}, removing the PPL module leads to a 3.0\% decrease in Pass Rate (from 42.0\% to 39.0\%) and an increase in the average steps required.
This degradation occurs because token-level pruning is agnostic to syntactic boundaries. In SE tasks, observations often consist of structured data such as JSON objects, stack traces, or code snippets. Randomly dropping tokens within these structures breaks the syntax, rendering the remaining context incomprehensible for the LLM. PPL-based segmentation ensures that we drop entire semantically coherent blocks (e.g., a redundant log line) rather than fragmenting critical code structures.

\begin{table}[h]
\caption{Ablation Study of Key Components. We compare the full \attncompress{} framework against variants where specific mechanisms are removed or replaced.}
\label{tab:ablation}
\resizebox{0.75\linewidth}{!}{%
\begin{tabular}{lcccccc}
\toprule
\textbf{Method} & \textbf{\textit{Pass (\%)}} & \textbf{\textit{Input (k)}} & \textbf{\textit{Output (k)}} & \textbf{\boldmath $C_{total}$ (\$)} & \textbf{\textit{Step}} & \textbf{\textit{PStep}} \\ \midrule
\textbf{\attncompress{} (Full)} & \textbf{42.0} & 547.27 & 10.74 & \textbf{0.0557} & \textbf{45.89} & 38.17 \\ \midrule
\textit{w/o PPL (Token-level)} & 39.0 & 622.37 & 10.88 & 0.0635 & 50.48 & 40.13 \\
\textit{w/o Attention (Random)} & 35.0 & 566.81 & \textbf{10.06} & 0.0567 & 47.38 & \textbf{35.89} \\
\textit{w/o Rolling (Fixed)} & 37.0 & \textbf{514.36} & 10.95 & 0.0619 & 48.44 & 39.51 \\ \bottomrule
\end{tabular}%
}
\end{table}

\paragraph{Analysis of Proxy Attention.}
To assess the impact of our attention score module, we replaced the proxy model's attention scoring with a random block selection strategy. This resulted in the most significant performance drop, with the pass rate plummeting to 35.0\% (-7.0\%).
This result underscores the high noise ratio in SE trajectories. The majority of tool outputs are irrelevant to the specific bug at hand. The attention mechanism acts as a critical filter, aligning the historical context with the agent's current reasoning. Without this guidance, the agent fails to locate the bug efficiently, forcing it to waste steps on unrelated files and often leading to task failure.

\paragraph{Analysis of Dynamic Rolling Window.}
We investigated the importance of dynamic context maintenance by disabling the rolling window mechanism (using a fixed strategy where context is compressed once and never revisited). The pass rate dropped significantly to 37.0\% (-5.0\%).
This validates the task focus drift hypothesis in debugging workflows. Information that appears irrelevant at step $t$ often becomes critical at step $t+20$ when the agent shifts its hypothesis. A static compression strategy permanently discards this information, preventing the agent from ``looking back''. The rolling window mechanism is therefore essential for allowing the agent to recall previously suppressed blocks as its focus evolves.

\mybox{Conclusion 2}{
The ablation study confirms that all three components are essential. The attention mechanism is the primary driver of effectiveness (+7.0\% pass rate) by filtering noise. The rolling window is crucial for handling non-linear debugging (+5.0\% pass rate), and PPL segmentation ensures syntactic integrity (+3.0\% pass rate) for code parsing.
}

\paragraph{Hyperparameter Sensitivity.}
We further examined the impact of five key hyperparameters on the validation set. The results are detailed in Table~\ref{tab:hyperparams}.

\begin{table}[h]
\caption{Hyperparameter Sensitivity Analysis on the Validation Set. Default settings are marked with *.}
\label{tab:hyperparams}
\resizebox{0.78\linewidth}{!}{%
\begin{tabular}{llcccccc}
\toprule
\textbf{Parameter} & \textbf{\textit{Value}} & \textbf{\textit{Pass (\%)}} & \textbf{\textit{Input (k)}} & \textbf{\textit{Output (k)}} & \textbf{\boldmath $C_{total}$ (\$)} & \textbf{\textit{Step}} & \textbf{\textit{PStep}} \\ \midrule
\multirow{3}{*}{\textbf{Tail Size ($k$)}} 
 & 2 (*) & 42.0 & 547.27 & 10.74 & 0.0557 & 45.89 & 38.17 \\
 & 5 & 42.0 & 563.19 & 10.69 & 0.0657 & 44.47 & 40.31 \\
 & 10 & \textbf{44.0} & 603.26 & 10.74 & 0.0821 & 43.64 & 38.77 \\ \midrule
\multirow{3}{*}{\textbf{Comp. Ratio ($\rho$)}} 
 & 0.1 & 40.0 & 465.49 & 10.17 & 0.0469 & 46.03 & 41.45 \\
 & 0.2 (*) & 42.0 & 547.27 & 10.74 & 0.0557 & 45.89 & 38.17 \\
 & 0.3 & \textbf{43.0} & 600.65 & 10.72 & 0.0613 & 45.39 & 41.35 \\ \midrule
\multirow{3}{*}{\textbf{Block Threshold}} 
 & -2 (*) & \textbf{42.0} & 547.27 & 10.74 & 0.0557 & 45.89 & 38.17 \\
 & 0 & \textbf{42.0} & 520.38 & 10.19 & 0.0519 & 45.23 & 35.69 \\
 & 2 & 39.0 & 516.52 & 10.41 & 0.0486 & 45.39 & 37.08 \\ \midrule
\multirow{4}{*}{\textbf{Proxy Layer}} 
 & Mean & 41.0 & 556.99 & 10.50 & 0.0541 & 46.58 & 40.90 \\
 & 0 (First) & 41.0 & 536.46 & 10.15 & 0.0532 & 46.00 & 35.88 \\
 & Middle & \textbf{43.0} & 548.67 & 10.32 & 0.0539 & 46.10 & 38.35 \\
 & -1 (Last) (*) & 42.0 & 547.27 & 10.74 & 0.0557 & 45.89 & 38.17 \\ \midrule
\multirow{3}{*}{\textbf{Window Size}} 
 & 2 & 35.0 & 541.79 & 10.79 & 0.0761 & 45.17 & 37.17 \\
 & 5 & 38.0 & 488.52 & 10.00 & 0.0529 & 43.11 & 35.32 \\
 & 10 (*) & \textbf{42.0} & 547.27 & 10.74 & 0.0557 & 45.89 & 38.17 \\ \bottomrule
\end{tabular}%
}
\end{table}

\textbf{Tail Size ($k$):} There is a trade-off between performance and cost. Increasing the tail size from 2 to 10 improves the pass rate to 44.0\%, as keeping more raw context helps the agent understand the immediate consequences of its actions. However, this comes at a significantly higher token cost (0.0821\$ vs 0.0557\$). To balance efficiency with effectiveness, and to maintain consistency with previous work like \agentdiet{}, we selected $k=2$ as the default.

\textbf{Compression Ratio ($\rho$):} As expected, a higher retention budget leads to better performance. Increasing $\rho$ to 0.3 yields a slight gain (43.0\%). However, reducing it to 0.1 causes a drop (40.0\%), indicating that essential information is being discarded. We selected $\rho=0.2$ as it offers a favorable cost-performance ratio, capturing the majority of relevant signals without inflating the context.

\textbf{Block Threshold:} The results favor smaller thresholds, which correspond to finer-grained segmentation. A threshold of -2 (adaptive) or 0 allows the model to select precise lines of interest, whereas a coarser threshold of 2 (grouping larger chunks) degrades performance to 39.0\%. Since the threshold has a negligible impact on computational overhead, we selected -2 to maximize segmentation flexibility.

\textbf{Proxy Layer selection:} The choice of proxy layer has only a marginal impact on the final performance, as the results across different layers (First, Middle, and Last) are largely comparable. This suggests that the relevance signal derived from proxy attention is robust. Given this insensitivity, we selected the Last layer (-1) as it is computationally free to extract and performs robustly.

\textbf{Window Size:} A larger rolling window is beneficial. Increasing the window size from 2 to 10 improves the pass rate from 35.0\% to 42.0\%. Meanwhile, increasing the Window Size has a minimal impact on token cost because the content within the window is already compressed. Therefore, we selected a larger window of $M=10$ to maximize the agent's ability to recall historical context.

\add{\note{AQ7} In summary, while \attncompress{} introduces several hyperparameters, Table~\ref{tab:hyperparams} demonstrates that they primarily govern cost-performance trade-offs rather than acting as fragile triggers. Key variables like compression ratio ($\rho$) and tail size ($k$) smoothly scale performance up or down, whereas architectural choices (threshold and proxy layer) show robust plateaus (e.g., both threshold -2 and 0 achieve 42\%, and middle/last layers achieve 43\% and 42\%). This indicates that the framework is resilient to parameter shifts and does not require exhaustive, task-specific fine-tuning.}

\mybox{Conclusion 3}{
\attncompress{} exhibits a clear trade-off between cost and performance. Notably, we adopted a conservative ``balanced'' configuration for our main evaluation rather than greedily maximizing the pass rate. This default setting already outperforms prior SOTA methods, suggesting that \attncompress{} is highly effective even under cost constraints, while offering significant headroom for performance improvement if larger token budgets are permitted.
}

\subsection{RQ3: Generalization Capabilities}

Finally, we assess the generalization capabilities of \attncompress{}. We investigate whether the framework maintains its effectiveness when utilizing different proxy model architectures and when applied to diverse programming languages beyond Python.

\paragraph{Proxy Model Robustness.}
\begin{table}[h]
    \centering
    \caption{\note{MR3, CQ1} \add{Proxy--agent attention alignment (\textit{Qwen3-Coder-30B} as reference scorer).}}
    \label{tab:proxy_attn_align}
    \resizebox{0.65\linewidth}{!}{%
    \begin{tabular}{lccc}
    \toprule
    \textbf{Proxy Model} & \textbf{Spearman $\rho$} & \textbf{Top-10\% Ovlp.} & \textbf{Top-20\% Ovlp.} \\ \midrule
    \textit{Qwen3-1.7B-Instruct} & 0.64 & 0.63 & 0.66 \\
    \textit{Qwen3-4B-Instruct} & 0.63 & 0.60 & 0.65 \\
    \textit{Qwen3-8B-Instruct} & 0.61 & 0.55 & 0.62 \\
    \textit{Gemma3-4B-Instruct} & 0.60 & 0.64 & 0.62 \\
    \textit{Llama3.2-3B-Instruct} & 0.56 & 0.63 & 0.63 \\ \bottomrule
    \end{tabular}%
    }
\end{table}

To verify that our approach is not dependent on a specific model family, we evaluated \attncompress{} using five distinct small language models (SLMs) as the proxy scorer on the 200-instance \swev{} test set. This includes three models from the \textit{Qwen3}~\cite{qwen3_4b_instruct_2507_modelcard} family to analyze scaling laws (1.7B, 4B, 8B), as well as \textit{Gemma3-4B-Instruct}~\cite{gemma_3_4b_it_modelcard} and \textit{Llama3.2-3B-Instruct}~\cite{llama_3_2_3b_instruct_modelcard} to test cross-architecture generalization.

\note{MR3, CQ1} \add{We first verify whether proxy relevance scoring aligns with the backend agent's needs. We use 200 uncompressed \original{} agent trajectories from the same 200-instance \swev{} subset evaluated below, and compare block-level attention rankings from each proxy model against \textit{Qwen3-Coder-30B} on identical PPL-segmented blocks. Table~\ref{tab:proxy_attn_align} reports spearman rank correlation ($\rho$), the fraction of shared blocks in the top 10\% of each ranking, and the same for the top 20\% (instance means). Our default \textit{Qwen3-4B-Instruct} proxy achieves $\rho{=}0.63$ with 0.60 and 0.65 top-10\%/20\% overlap, indicating substantial agreement on the blocks most likely to be retained under our compression budget. Cross-family proxies remain in a similar range, suggesting that imperfect global rankings still preserve the salient context the agent attends to.}

\begin{table*}[t]
\caption{\note{MR3, CQ1, CQ2}\add{Performance and Efficiency with different Proxy Models.}}
\label{tab:proxy_generalization}
\resizebox{0.85\textwidth}{!}{%
\begin{tabular}{llccccc}
\toprule
\textbf{Agent LLM} & \textbf{Proxy Model} & \textbf{\textit{Pass (\%)}} & \textbf{\textit{Input (k)}} & \textbf{\boldmath $C_{total}$ (\$)} & \textbf{\textit{Step}} & \textbf{\textit{Ana Time (s)}} \\ \midrule
\multirow{5}{*}{\textit{Qwen3-Coder-30B}}
& \textit{Qwen3-1.7B-Instruct} & 43.50 & 551.63 & 0.0571 & 47.16 & \textbf{71.61} \\
& \textit{Qwen3-4B-Instruct} & 44.50 & 534.62 & 0.0551 & 46.30 & 530.87 \\
& \textit{Qwen3-8B-Instruct} & \textbf{46.50} & 558.34 & 0.0607 & 46.67 & 384.13 \\
& \textit{Gemma3-4B-Instruct} & 45.00 & 694.98 & 0.0588 & 55.02 & 515.63 \\
& \textit{Llama3.2-3B-Instruct} & 45.50 & \textbf{507.71} & \textbf{0.0548} & \textbf{44.98} & 100.37 \\ \midrule
\multirow{5}{*}{\textit{Gemini-3-Flash}}
& \textit{Qwen3-1.7B-Instruct} & 69.00 & \textbf{749.32} & 0.1940 & 64.90 & \textbf{131.49} \\
& \textit{Qwen3-4B-Instruct} & 71.50 & 780.17 & \textbf{0.1570} & \textbf{60.46} & 252.47 \\
& \textit{Qwen3-8B-Instruct} & \textbf{72.50} & 773.15 & 0.2077 & 65.78 & 682.06 \\
& \textit{Gemma3-4B-Instruct} & 68.50 & 781.90 & 0.1705 & 71.48 & 790.62 \\
& \textit{Llama3.2-3B-Instruct} & 71.50 & 781.39 & 0.2072 & 68.41 & 239.08 \\ \bottomrule
\end{tabular}%
}
\end{table*}

The end-to-end results are presented in Table~\ref{tab:proxy_generalization}, where we report both \textit{Qwen3-Coder-30B} and \textit{Gemini-3-Flash} as backend agents. \note{MR3, CQ1} \add{Together with Table~\ref{tab:proxy_attn_align}, this shows that small--big LLM attention agreement is sufficient for stable agent behavior: pass rates vary by only a few points across proxies.}

The data indicates a high degree of transferability across model families. On \textit{Qwen3-Coder-30B}, both \textit{Gemma3-4B-Instruct} (45.0\%) and \textit{Llama3.2-3B-Instruct} (45.5\%) achieve competitive pass rates comparable to \textit{Qwen3-4B-Instruct} (44.5\%). \note{MR3, CQ1, CQ2} \add{The same trend holds on \textit{Gemini-3-Flash}, where \textit{Gemma3-4B-Instruct} and \textit{Llama3.2-3B-Instruct} reach 68.5\% and 71.5\%, respectively, close to the default \textit{Qwen3-4B-Instruct} (71.5\%). This is consistent with their strong top-10\%/20\% overlap with the backend agent in Table~\ref{tab:proxy_attn_align} (e.g., 0.63, 0.64 for \textit{Llama3.2-3B-Instruct}),} suggesting that the attention patterns used to distinguish signal from noise are universal features present in various LLM families, making \attncompress{} model-agnostic.

We further analyze the impact of model size within the \textit{Qwen3} series using \textit{Qwen3-Coder-30B} as the backend agent. There is a nuanced trade-off between proxy model size, selection quality, and compression latency (\textit{Ana Time} column). Specifically, \textit{Qwen3-1.7B-Instruct} is the fastest (71.6s) but suffers from a slight performance drop (43.5\%), suggesting that very small models may struggle to accurately identify subtle long-term dependencies beyond coarse attention agreement. \textit{Qwen3-8B-Instruct} achieves the highest pass rate of 46.5\%. However, this gain comes at a higher analysis time (384.1s). Therefore, \textit{Qwen3-4B-Instruct} strikes the optimal balance. It improves pass rate over the 1.7B model (+1.0\%) while incurring substantially lower analysis time than the 8B proxy. 

\note{MR3, CQ1, CQ2} \add{We also observe that the ranking of proxy models based on end-to-end agent performance (Table~\ref{tab:proxy_generalization}) does not strictly align with their attention alignment scores (Table~\ref{tab:proxy_attn_align}). For instance, \textit{Qwen3-1.7B-Instruct} achieves the highest top-20\% overlap (0.66) but records the lowest pass rate on \textit{Qwen3-Coder-30B}. We attribute this discrepancy primarily to two factors: (i) the differences in alignment across proxies are marginal (all top-20\% overlaps fall within $[0.62, 0.66]$), meaning that minor variations in ranking do not necessarily translate to downstream success; and (ii) end-to-end evaluations inherently introduce variance due to agent stochasticity. While the exact mechanism by which attention alignment influences final agent outcomes warrants further investigation, our results yield two practical takeaways: proxy models from diverse families and scales share broadly similar attention distributions, and this shared structure enables effective compression with negligible variations in pass rates across proxies.}

\mybox{Conclusion 4}{
The effectiveness of attention-based filtering is not tied to a single architecture; both \textit{Gemma} and \textit{Llama} models provide sufficient semantic signals to drive high agent performance.
}

\paragraph{Multi-Language Generalization.}
We extended our evaluation to \mswef{} to test adaptability across seven programming languages: C, C++, Go, Java, JavaScript, Rust, and TypeScript. Table~\ref{tab:multi_lang_main} summarizes the overall results, and Figure~\ref{fig:multi_lang} illustrates the specific pass counts per language.

\begin{table}[ht]
\centering
\begin{minipage}{0.62\linewidth}
\centering
\captionof{table}{Main Results on \mswef{}. \textbf{Bold} indicates the best performance among compression methods; \underline{Underline} indicates the second best.}
\label{tab:multi_lang_main}
\resizebox{\linewidth}{!}{%
\begin{tabular}{lcccccc}
\toprule
\textbf{Method} & \textbf{\textit{Pass (\%)}} & \textbf{\textit{Input (k)}} & \textbf{\textit{Output (k)}} & \textbf{\boldmath $C_{total}$ (\$)} & \textbf{\textit{Step}} & \textbf{\textit{PStep}} \\ \midrule
\original{} & 20.33 & 1629.38 & 11.69 & 0.1057 & 52.83 & 42.66 \\ \midrule
\random{} & 17.00 & 1420.96 & 14.54 & 0.1074 & 79.51 & 62.24 \\
\lingua{} & 16.00 & 1360.88 & 16.50 & 0.1046 & 78.24 & 58.69 \\
\summary{} & 17.39 & 1107.88 & \textbf{11.54} & 0.0971 & \underline{59.07} & \underline{45.62} \\
\obsmask{} & 15.72 & \textbf{719.77} & 12.35 & \textbf{0.0376} & 72.05 & 55.85 \\
\add{\sliding{}} & 17.67 & \underline{790.56} & 12.02 & \underline{0.0764} & 61.76 & 52.36 \\
\agentdiet{} & \underline{18.33} & 1014.70 & \underline{11.56} & 0.1132 & \textbf{58.45} & \textbf{45.04} \\ \midrule
\textbf{\attncompress{}} & \textbf{19.67} & 879.77 & 11.90 & 0.0826 & 62.38 & 49.32 \\ \bottomrule
\end{tabular}%
}
\end{minipage}
\hfill
\begin{minipage}{0.36\linewidth}
\centering
\includegraphics[width=\linewidth]{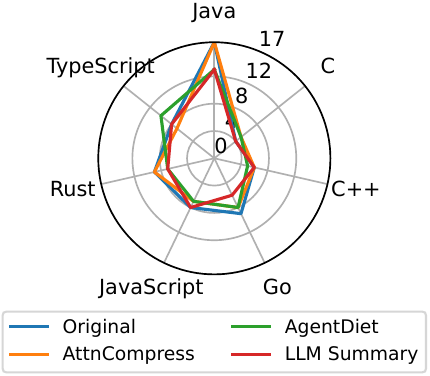}
\captionof{figure}{Number of passed instances per language on \mswef{}. \attncompress{} matches the \original{} baseline in Java, Rust, and C++, outperforming other compression methods.}
\label{fig:multi_lang}
\end{minipage}
\end{table}

\attncompress{} achieves the best performance among all compression methods, with a pass rate of 19.67\%, which is close to the \original{} full-context baseline (20.33\%). In contrast, other methods show significant degradation: \obsmask{} drops to 15.72\%, \sliding{} achieves 17.67\%, and \agentdiet{} achieves 18.33\% while incurring higher costs (0.1132\$ vs 0.0826\$).

Language-specific analysis (Figure~\ref{fig:multi_lang}) reveals that \attncompress{} maintains parity with the \original{} baseline in strictly typed and verbose languages.
\begin{itemize}
\item \textbf{C/C++ \& Java:} In Java, \attncompress{} matches the \original{} baseline exactly (17 passed) and outperforms \agentdiet{} (13 passed). Similarly, in C, it successfully solves 5 instances compared to 4 for the \original{} baseline. This indicates that our method remains effective in verbose, statically-typed environments.
\item \textbf{Rust \& Go:} Performance is stable in modern systems languages. In Rust, \attncompress{} matches the \original{} baseline (9 passed), and in Go, it shows a marginal difference (8 vs. 9).
\item \textbf{Web Languages (TS/JS):} For TypeScript and JavaScript, \attncompress{} (7 passed) performs slightly below the \original{} (8 passed) but remains competitive with \agentdiet{}.
\end{itemize}

\mybox{Conclusion 5}{
\attncompress{} generalizes across diverse programming languages, achieving near-parity with full-context agents (96.7\% relative performance) while reducing token costs by 20\%.
}

\section{Threats to Validity}
\label{sec:threats}

\noindent\textbf{Threats to Internal Validity.}
The primary internal threat is Data Leakage, as proprietary LLMs might have seen the \swebench{} issues during training. We mitigated this by including the more recent \mswebench{} in our evaluation. Furthermore, since all baselines utilize the same backend models, any leakage affects them equally, preserving the validity of relative comparisons.
A second threat is hyperparameter overfitting. To address this, we strictly isolated a 100-instance validation set for ablation studies and parameter tuning, ensuring the reported performance on the 200-instance test set reflects genuine generalization rather than overfitting.

\noindent\textbf{Threats to External Validity.}
The main external threat is generalization across agent frameworks. Due to computational costs, we evaluated \attncompress{} primarily on \texttt{Trae-Agent}. However, as most SE agents follow similar ReAct patterns, our middleware approach is theoretically transferable.
We also addressed model and language generalization by validating our framework across diverse proxy models (\textit{Qwen}, \textit{Llama}, \textit{Gemma}) and seven programming languages. The consistent results suggest our findings are not limited to a specific model architecture or language ecosystem.

\noindent\textbf{Threats to Construct Validity.}
A threat to construct validity is the reliance on test-based evaluation. A patch passing available tests (plausible) may not be semantically equivalent to the developer's fix (correct). While this is a known limitation of benchmarks like \swebench{}, it serves as a standard proxy for task success. Crucially, this metric limits all comparison methods equally; therefore, the observed improvements in pass rate reliably indicate that \attncompress{} retains more critical task information than other compression baselines.

\section{Conclusions}
\label{sec:conclusions}
In this paper, we addressed the critical context scalability bottleneck in Autonomous Software Engineering (ASE) agents. We introduced \attncompress{}, a dynamic compression framework that overcomes the limitations of static pruning and heuristic summarization through three key mechanisms: structure-aware segmentation via PPL spikes, proxy attention-guided relevance estimation, and a dynamic rolling window. This approach ensures the preservation of syntactic integrity and semantic dependencies essential for SE tasks. Extensive evaluation on \swev{} demonstrates that \attncompress{} achieves a state-of-the-art pass rate of 53.17\%, outperforming strong compression baselines while reducing token consumption by over 21.6\% and total costs by 33.6\%. Our results confirm that dynamic attention alignment offers a superior, model-agnostic solution for efficient, long-horizon software engineering tasks.

\section*{Data Availability}
The replication package for our study, containing the necessary source code and scripts to reproduce our experiments, is available at the repository~\cite{AttnCompress}.

\newpage{}

\bibliographystyle{ACM-Reference-Format}
\bibliography{ref}

\end{document}